\documentclass[sigconf]{acmart}

\usepackage{booktabs} % For formal tables

\usepackage{amsmath}
\usepackage{graphicx}
\usepackage{enumerate}
\usepackage{color}
\usepackage{mathtools}
\usepackage{algorithm}
\usepackage{algpseudocode}
\usepackage{balance}
\usepackage{makecell}
\copyrightyear{2018} 
\acmYear{2018} 
\setcopyright{acmcopyright}
\acmConference{2018 KDD Workshop on Machine Learning for Medicine and Healthcare}{}{Aug. 19-- Aug 24, 2018, London, U.K.}

\begin{document}
\title[PGLasso]{PGLasso: Microbial Community Detection through Phylogenetic Graphical Lasso}
%\title[PGLasso]{Automated and accurate microbial community identification using phylogenetic information}
%\titlenote{Produces the permission block, and copyright information}
%\subtitle{MTPLasso}
%\subtitlenote{The full version of the author's guide is available as
 % \texttt{acmart.pdf} document}

\author{Chieh Lo}
%\authornote{Dr.~Trovato insisted his name be first.}
\affiliation{%
  \institution{Carnegie Mellon University}
  \city{Pittsburgh} 
  \state{PA} 
  \postcode{15213-3890}
}
\email{chiehl@andrew.cmu.edu}

\author{Radu Marculescu}
%\authornote{The secretary disavows any knowledge of this author's actions.}
\affiliation{%
  \institution{Carnegie Mellon University}
    \city{Pittsburgh} 
  \state{PA} 
  \postcode{15213-3890}
}
\email{radum@cmu.edu}


\begin{abstract}
Due to the recent advances in high-throughput sequencing technologies, it becomes possible to directly analyze microbial communities in the human body and in the environment. Knowledge of how microbes interact with each other and form functional communities can provide a solid foundation to understand microbiome related diseases; this can serve as a key step towards precision medicine. In order to understand how microbes form communities, we propose a two step approach: First, we infer the microbial co-occurrence network by integrating a graph inference algorithm with phylogenetic information obtained directly from metagenomic data. Next, we utilize a network-based community detection algorithm to cluster microbes into functional groups where microbes in each group are highly correlated. We also curate a ``gold standard" network based on the microbe-metabolic relationships which are extracted directly from the metagenomic data. Utilizing community detection on the resulting microbial metabolic pathway bipartite graph, the community membership for each microbe can be viewed as the true label when evaluating against other existing methods. Overall, our proposed framework Phylogenetic Graphical Lasso (PGLasso) outperforms existing methods with gains larger than 100\% in terms of Adjusted Rand Index (ARI) which is commonly used to quantify the goodness of clusterings. 
\end{abstract}




\maketitle

\section{Introduction}
Microbes play an important role in human life. Additionally, microbial communities (\textit{i.e.,} groups of microbes) can exhibit a rich dynamics including the way they adapt, develop, and interact with the human body and the surrounding environment. However, the way microbes affect the human health and how they form communities remains largely unknown. Knowledge of how microbes interact with each other and form functional communities can provide a solid foundation to understand microbiome related diseases; this can serve as a key step towards precision medicine~\cite{Lu2014}. 

To uncover how microbes form functional communities that relate to certain diseases, one way is to utilize unsupervised clustering methods to group microbes that have similar functionalities together. Recently, metagenomic data provided rich information about human microbiome including microbial abundance at different body sites and microbe-metabolic pathway relationships. However, there exist two main challenges that stem from the very nature of the metagenomic sequencing data that may result in inaccurate information: First, the publicly available data only contain a few hundreds of samples ($n$), while the number of measured microbes ($p$) usually ranges from hundreds to thousands; therefore, the number of associations to be inferred can possibly get as high as $p(p-1)/2$; this results in high-dimensional data since $p(p-1)/2 \gg n$. Second, metagenomic data can only provide the \textit{relative} abundance of various species; this is because the sequencing results are a function of sequencing depth and the biological sample size~\cite{Sims2014}. Therefore, from a statistical standpoint, the \textit{relative} taxa abundance falls into the class of compositional data~\cite{Aitchison1986}; this can greatly affect the performance of existing machine learning algorithms. Consequently, traditional unsupervised clustering methods such as $k$-means and Gaussian mixture model~\cite{bishop2006} exhibit two main drawbacks: First, the number of clusters needs to be setup before running those algorithms. Second, features of metagnenomic data such as microbial abundance are generally sparse and compositional in nature. 

In this paper, we propose a new method which builds a network model, where nodes and edges represent microbes and their relationships (\textit{e.g.}, associations or interactions), respectively. This representation has the following advantages over traditional unsupervised clustering methods: First, we can directly use network-based community detection algorithms to circumvent the problem of choosing the number of clusters. Second, we can incorporate known information to mitigate the problem of high-dimensionality and compositionality while learning the network structure. Third, the network can provide much more insight (compared to traditional clustering methods) such as identify which microbes are the ``key players" thus greatly affecting the other microbes and their growth. 

\begin{figure*}[!t]%figure2
\centering
\includegraphics[width=0.9\linewidth]{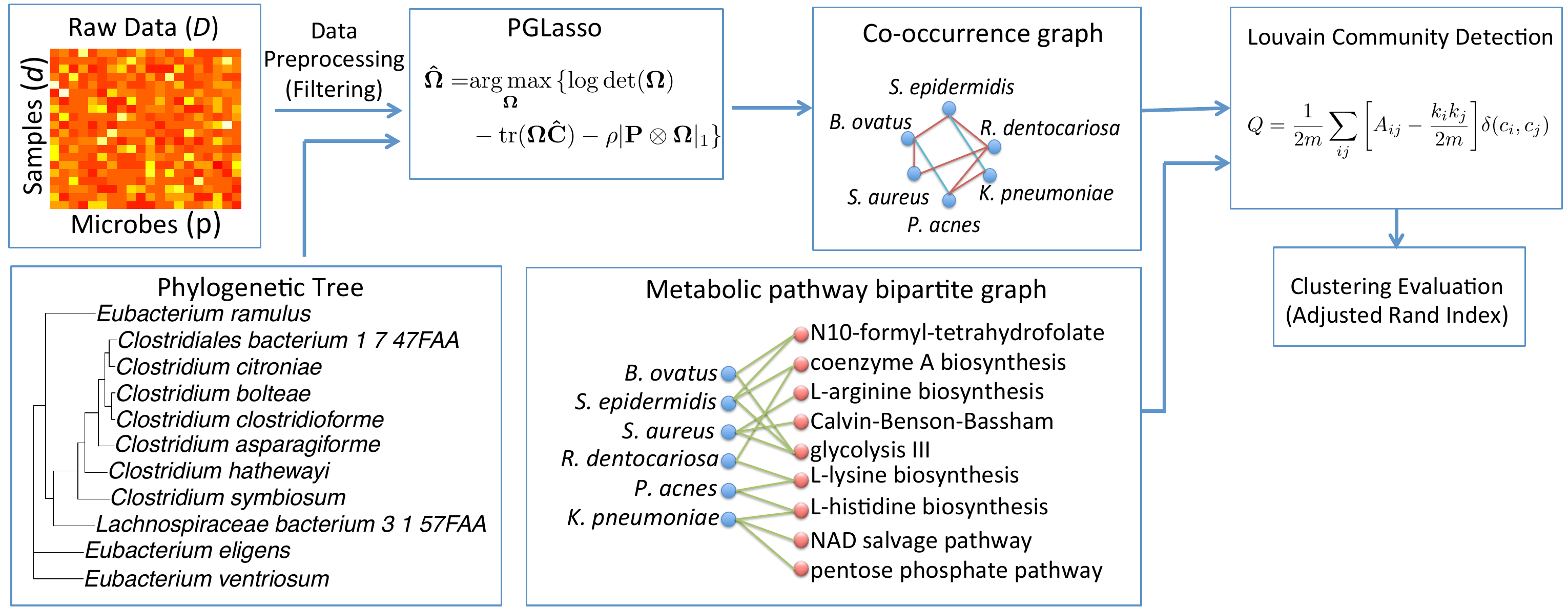}
\caption{The proposed PGLasso pipeline. We first pre-process the microbial metagenomic data and integrate the phylogenetic information. We then use our proposed PGLasso to infer the microbial co-occurrence network. Next, we conduct Louvain community detection algorithm on the inferred co-occurrence network and the curated metabolic pathway bipartite network. Finally, we use Adjusted Rand Index (ARI) to quantify the clustering results.}\label{fig:system}
\vspace{-10pt}
\end{figure*}

As shown in Fig.~\ref{fig:system}, our proposed framework (PGLasso) consists of a graph inference algorithm and a community detection method. For the graph inference part, we propose to integrate knowledge from the phylogenetic information created from PhyloPhlAn~\cite{Segata2013} into a graph inference algorithm; this can mitigate both high-dimensionality and compositionality problems that result from the very nature of the metagenomic data. The rationale behind the incorporation of phylogenetic information concerns the assortativity that has been widely observed in other microbial network studies~\cite{Faust2012}.  Next, based on the inferred microbial co-occurrence (MC) network, we can directly apply a network-based community detection algorithms such as the Louvain community detection algorithm~\cite{Blondel2008} to maximize the modularity of the MC network; the resulting grouping of microbes is highly correlated and may correspond to certain disease phenotype.

However, there exists neither a true correlation network of microbe-microbe associations, nor a best clustering of microbes in real experimental data. In order to assess and compare the performance of clustering on the network inferred with different methods, we curate a ``gold standard" dataset based on the microbe-metabolic pathway relationships generated from the Humann2 pipeline~\cite{Shar2012}; these relationships can be viewed as a bipartite network where node and edge sets ($V$ and $E$) correspond to microbes and metabolic pathways, respectively. An edge is connected between two nodes if there exists a metabolic pathway in the microbes. By performing clustering on the bipartite network via maximizing the modularity, the resulting community membership for each microbe can be viewed as the \textit{true} label when evaluating different methods. 

Finally, to evaluate the goodness of microbiome community detection\footnote{In this paper, we use the terms community detection and clustering interchangeably.}, we utilize the Adjusted Rand Index~\cite{Hubert1985} which is commonly used to quantify the performance of unsupervised clustering. However, the quantitative value of most metrics are task dependent.  In order to provide a basis for evaluating different methods, we simulate a null model by randomly re-routing the inferred microbial network. Evaluation results on null models can be seen as lower bounds for each metric. We compare our clustering results against other existing algorithms and the null model and show that our proposed PGLasso achieves the best clustering results with a huge gain ($>100\%$).

\section{Methods}
\subsection{Acquisition and preprocessing of metagenomic data}\label{sec:data}
In this paper, we consider high-throughput comparative metagenomic data generated by the next-generation sequencing (NGS) platforms. Data obtained from the human microbiome project (HMP) have a curated collection of  microorganisms sequence associated with the human body from shotgun sequencing technologies. We obtain data from \url{http://hmpdacc.org/HMASM/} and use the trimmed sequences as inputs to the Humann2~\cite{Shar2012} pipeline which can generate the microbial abundance table and microbe-metabolic pathway relationship for each sample.

The resulting microbial abundance table can be represented by a matrix $D \in \mathbb{N}^{n\times p}$ where $\mathbb{N}$ represents the set of natural numbers. $d^{i} = [d_1^{i}, d_2^{i}, \dots, d_p^{i}]$ denotes the $p$-dimensional row vector of relative taxonomy abundance from the $i^{th}$ sample ($i = 1, \dots, n$). We then use the log-ratio transform (\citealp{Aitchison1986}) to account for different depths of sequencing samples. Statistical inference on the log-ratio transform of the compositional data ($x$) can be shown to be equivalent to the log-ratio transform on the unobserved absolute abundance ($d$): $\log(\frac{x_i}{x_j}) = \log(\frac{d_i/m}{d_j/m}) = \log(\frac{d_i}{d_j}).$  Here, we apply the centered log-ratio (clr) transform as: $c = \text{clr}(x) = [\log(\frac{x_1}{m(x)}), \log(\frac{x_2}{m(x)}), \dots, \log(\frac{x_p}{m(x)})]$, where $m(x) = (\prod_{i=1}^{p} x_i)^{\frac{1}{p}}$ is the geometric mean of the composition vector $x$. The resulting vector $c$ is constrained to be a zero sum vector.
%\begin{align}
%c = \text{clr}(x) = [\log(\frac{x_1}{m(x)}), \log(\frac{x_2}{m(x)}), \dots, \log(\frac{x_p}{m(x)})] \label{eq:clr}
%\end{align}
%where $m(x) = (\prod_{i=1}^{p} x_i)^{\frac{1}{p}}$ is the geometric mean of the composition vector $x$. The resulting vector $c$ is constrained to be a zero sum vector.

\begin{table}[!th]
\centering
\small
\begin{tabular}{c|cccc}\hline
Dataset&		($n$, $p$)&	\#Pathways&	\#MC&	\#MPB \\ \hline
AntNar&		(91, 13)&		72&			49&	133\\
BucMuc&		(113, 71)&		185&			560&	2452\\
SupPa&		(124, 129)&	194&			995&	3508	\\
Stool&			(143, 83)&		256&			817&	2009	\\
TonDor&		(130, 103)&	192&			664&	2751\\ \hline
\end{tabular}
\caption{Statistics summary. $n$ and $p$ represent the number of sample and microbes, respectively.  \#Pathways, \#MC, and \#MPB represents the number of pathways found, the number of inferred edges in microbial co-occurrence network, and the number of edges in metabolic pathway bipartite network, respectively. Abbreviations: AntNar: Anterior nares, BucMuc: Buccal mucosa, SupPla: Supragingival plague, TonDor: Tongue dorsum.}
\label{Tab:stats}
\vspace{-20pt}
\end{table} 

\subsection{Microbial co-occurrence (MC) network}\label{sec:MC}
\subsubsection{Graph inference algorithm}
To infer the pairwise associations among microbes, we can transform the original inferring problem into a graph inference problem where each node represents a microbe and each edge represents a pairwise association between microbes; the resulting network is an undirected graph $\mathcal{G} = (V, E)$, where $V$ and $E$ represent the node and edge sets, respectively. 

Suppose the observed data ($d$) are drawn from a multivariate normal distribution $N(d|\mu, \boldsymbol{\Sigma})$ with mean $\mu$ and covariance $\boldsymbol{\Sigma}$. The inverse covariance matrix (precision matrix) $\boldsymbol{\Omega} = \boldsymbol{\Sigma^{-1}}$ encodes the conditional independence among nodes. More specifically, if the entry ($i, j$) of the precision matrix $\boldsymbol{\Omega}_{i,j} = 0$, then node $i$ and node $j$ are conditionally independent (given the other nodes) and there is no edge among them (i.e., $E_{i,j} = 0$).  One suitable algorithm to select the precision matrix under sparsity assumption is to utilize the Graphical Lasso (GLasso) proposed previously in~\cite{Friedman2008, Kurtz2015}.

%However, microbial data usually come with a finite amount of samples ($n$) but with high dimensionality ($p$); this makes the graph inference problem intractable since the number of variables ($\frac{p(p-1)}{2}$) is greater than $n$. To solve this problem, an important assumption that needs to be made is to assume that the underlying (true) network is reasonably sparse. One suitable algorithm to select the precision matrix under sparsity assumption is to utilize the graphical Lasso proposed previously in~\cite{Friedman2008, Kurtz2015}.

\subsubsection{Prior information: phylogenetic tree}\label{sec:tree}
Phylogenetic relationships among microbes serve as an important information of how closely microbes relate to each other from an evolution aspects. Several studies have shown that phylogenetically correlated microbes are more likely to interact to each other~\cite{Faust2012}. One way to quickly access the pairwise relationship among microbes is to build a phylogenetic tree based on the protein sequences of the microbial genomes. The phylogenetic tree\footnote{A leaf of the resulting tree represents a microbe.} is built using \textit{de novo} methods by using the automated tool PhyloPhlAn~\cite{Segata2013}. Next, we utilize the cophenetic distances~\cite{ape2004} to compute the pairwise distance among all pairs of microbes presented in the dataset. 

Given the phylogenetic relationships, we can obtain the prior information and represent it as a matrix $\mathbf{P} \in \mathbb{R}^{p \times p}$, where each entry $\mathbf{P}_{i, j}\in [0,1]$ represents the prior probability of associations between microbe $i$ and microbe $j$. We can impose different amounts of penalties on the precision matrix; this is different from the standard formulation of GLasso where the penalty ($\rho$) {\color{black}{imposed}} on the precision matrix is the same. Therefore, by incorporating the prior information into the penalty matrix ($\mathbf{P}$), the proposed PGLasso can be formulated as follows: 
\begin{align}
\boldsymbol{\hat{\Omega}} = {\displaystyle {\underset {\boldsymbol{\Omega}}{\operatorname {arg\,max} }}\, \{ \log \text{det} (\boldsymbol{\Omega}) - \text{tr} (\boldsymbol{\Omega} \mathbf{\hat{C}}) - \rho | \mathbf{P} \otimes \boldsymbol{\Omega}|_1} \}
\end{align}
where $\mathbf{\hat{C}}$ is the empirical covariance of the microbial abundance data, and $\boldsymbol{\Omega}$ is the precision matrix of the estimated associations among microbes. Here, det and tr denote the determinant and the trace of a matrix, respectively. $|\boldsymbol{\Omega}|_1$ is the $L_1$ norm, i.e., the sum of the absolute values of the elements of $\boldsymbol{\Omega}$ and $\otimes$ represents the component-wise multiplication. When the value of $\mathbf{P}_{i,j}$ is large, this directly imposes a heavy penalty and represents a weaker association between taxa and vice versa. This way, by imposing the phylogenetic information, we can accurately infer the associations among microbes.  

\begin{figure}[!t]%figure2
\centering
\includegraphics[width=0.9\linewidth]{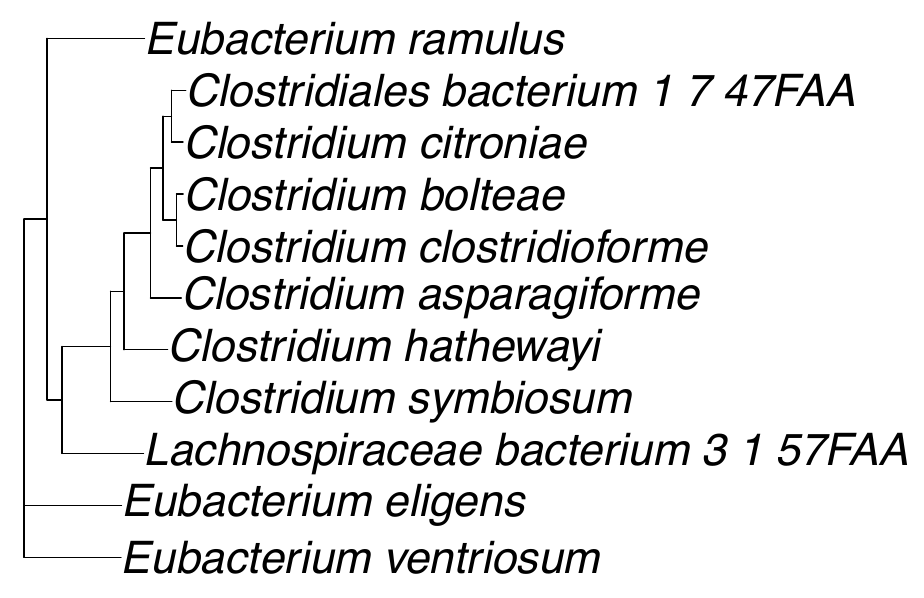}
\caption{Visualization of a phylogenetic tree obtained from the HMP HMASM dataset. As it can be seen, microbes in the same genus group are close to each other within the phylogenetic tree.   
}\label{fig:tree}
\vspace{-10pt}
\end{figure}

\subsection{Metabolic pathway bipartite (MPB) network}\label{sec:MPB}
We curate ``gold standard" datasets based on the microbe-metabolic pathway relationships. The rationale to curate such datasets lies in the fact that microbes that share and exchange same metabolites are more likely to interact. For instance, microbes that compete for certain metabolites can have negative interactions among them. This type of behavior has been observed in existing literature and has been utilized to construct pair-wise metabolic models of microbes~\cite{Mendes2016}. 

We can compute the relationships of microbe-metabolic pathway by utilizing the Humann2 pipeline~\cite{Shar2012}. The obtained relationships can be viewed as a bipartite network where node and edge sets $V$ and $U$ correspond to microbes and metabolic pathways, respectively. An edge is inserted between two sets of nodes if a certain pathway exists in the microbe. By performing community detection on the bipartite network with the objective of maximizing the modularity, the resulting community membership (\textit{i.e.,} clustering) for each microbe can be viewed as the \textit{true} label when evaluating different methods. 

\subsection{Louvain community detection algorithm}\label{sec:Lu}
The best possible clustering of the nodes of a given network can be detected by maximizing a modularity quality function. In this paper, we optimize the modularity of a given network by utilizing the Louvain method for community detection. The modularity function ($Q$) to be optimized is defined as follows:
\begin{align}
Q = {\frac {1}{2m}}\sum \limits _{ij}{\bigg [}A_{ij}-{\frac {k_{i}k_{j}}{2m}}{\bigg ]}\delta (c_{i},c_{j})
\end{align}
where $A_{ij}$ represents the edge weight between nodes $i$ and $j$. $k_{i}$ and $k_j$ are the sum of the weights of the edges attached to nodes $i$ and $j$, respectively; $2m$ is the sum of all of the edge weights in the network; $c_{i}$ and $c_{j}$ are the communities of the nodes belong to; and $\delta$ represents the delta function.

\begin{table}[!t]
\centering
\small
\begin{tabular}{c|cccc}\hline
Dataset&		PGLasso&			GLasso&			SparCC&			Random Null \\ \hline
AntNar&		0.303 (0.090)&	0.289 (0.048)&	0.099 (0.087)&	0.056 (0.051)\\
BucMuc&		0.194 (0.075)&	0.022 (0.017)&	0.016 (0.012)&	0.019 (0.011)	\\
SupPa&		0.201 (0.047)&	0.016 (0.016)&	0.009 (0.005)&	0.007 (0.005)	\\
Stool&			0.132 (0.028)&	0.018 (0.016)&	0.012 (0.010)&	0.017 (0.011)	\\
TonDor&		0.179 (0.043)&	0.033 (0.020)&	0.011 (0.010)&		0.010 (0.009)	\\ \hline
\end{tabular}
\caption{Performance comparison of PGLasso, GLasso, SparCC and the random null models using Adjusted Rand Index. We average over 20 runs of community detection with standard deviations shown in round brackets. Abbreviations: AntNar: Anterior nares, BucMuc: Buccal mucosa, SupPla: Supragingival plague, TonDor: Tongue dorsum. }
\label{Tab:performances}
\vspace{-20pt}
\end{table}

\subsection{Evaluation metrics}
The Rand Index (RI) is defined as a similarity measure between two clusterings by considering all pairs of samples in the predicted and true clusterings. Mathematically speaking, given two different clusterings of microbes, namely $X = \{ X_1, X_2, \ldots , X_r \}$ and $Y = \{ Y_1, Y_2, \ldots , Y_s \}$, where $X$ and $Y$ are the clustering results of the inferred MC network and the MPB network, the number of overlapping microbes in $X_{i}$ and $Y_{j}$ is then used to compute the RI. %  $X_i$ and  $|X_{i}\cap Y_{j}|$ between $X$ and $Y$ can be summarized in a table $N$ where each entry $n_{ij}=|X_{i}\cap Y_{j}|$. We can then use $N$ to compute the Rand index.  

%The raw RI score is then ?adjusted for chance? into the ARI score using the following scheme:

We use the Adjusted Rand Index (ARI) to evaluate the clustering results of our inferred MC network against the MPB network. The ARI is the corrected-for-chance version of the RI and is defined as: $ARI = \frac{(RI - \mathbf{E}[RI])}{(\max(RI) - \mathbf{E}[RI])}$, where $\mathbf{E}$ represent the expected RI used to adjust for random chance. 

%Given $p$ microbes, and two different clusterings of these elements, namely  $X = \{ X_1, X_2, \ldots , X_r \}$ and $Y = \{ Y_1, Y_2, \ldots , Y_s \}$, where $X$ and $Y$ are the clustering results of the inferred MC network and the MPB network. 
%
%The overlap between $X$ and $Y$ can be summarized in a contingency table $N$ where each entry $n_{ij}=|X_{i}\cap Y_{j}|$. We can then calculate the adjusted Rand index by computing the Rand index minus the expected Rand index. 

%${\displaystyle \overbrace {ARI} ^{\text{Adjusted Index}}={\frac {\overbrace {\sum _{ij}{\binom {n_{ij}}{2}}} ^{\text{Index}}-\overbrace {[\sum _{i}{\binom {a_{i}}{2}}\sum _{j}{\binom {b_{j}}{2}}]/{\binom {n}{2}}} ^{\text{Expected Index}}}{\underbrace {{\frac {1}{2}}[\sum _{i}{\binom {a_{i}}{2}}+\sum _{j}{\binom {b_{j}}{2}}]} _{\text{Max Index}}-\underbrace {[\sum _{i}{\binom {a_{i}}{2}}\sum _{j}{\binom {b_{j}}{2}}]/{\binom {n}{2}}} _{\text{Expected Index}}}}} {\displaystyle \overbrace {ARI} ^{\text{Adjusted Index}}={\frac {\overbrace {\sum _{ij}{\binom {n_{ij}}{2}}} ^{\text{Index}}-\overbrace {[\sum _{i}{\binom {a_{i}}{2}}\sum _{j}{\binom {b_{j}}{2}}]/{\binom {n}{2}}} ^{\text{Expected Index}}}{\underbrace {{\frac {1}{2}}[\sum _{i}{\binom {a_{i}}{2}}+\sum _{j}{\binom {b_{j}}{2}}]} _{\text{Max Index}}-\underbrace {[\sum _{i}{\binom {a_{i}}{2}}\sum _{j}{\binom {b_{j}}{2}}]/{\binom {n}{2}}} _{\text{Expected Index}}}}}$

\section{Experimental Results}\label{syn_exp}
\subsection{Network inference and clustering}
We consider five different body sites from the HMP HMASM dataset\footnote{https://www.hmpdacc.org/HMASM/} as shown in Table~\ref{Tab:stats}. More specifically, we first pre-process the metagenomic sequencing data to detect the presence of microbes and the microbe-metabolic pathway relationships as described in Section~\ref{sec:data}. Next, we collect the genome sequence for each microbe from PATRIC~\cite{Wattam2017} and compute the phylogenetic trees as described in Section~\ref{sec:tree}. Third, we run our proposed inference algorithm PGLasso to obtain the MC network. Finally, we conduct the Louvain community detection algorithm introduced in Section~\ref{sec:Lu} to find the best clustering of MPB network and the MC network. 

\subsection{Evaluation on clustering quality}
To show the applicability of our proposed PGLasso, we consider and compare with a well-known algorithm, SparCC~\cite{Friedman2012},  which is mainly developed for inferring the MC networks. Other than that, we consider the GLasso algorithm without using any prior knowledge and also generate a random network by randomly re-routing the edges inferred by the PGLasso; this serves as a null model to quantify the relative gains of different methods.  

As can be seen in Table~\ref{Tab:performances}, PGLasso outperforms other methods in terms of the ARI; this shows that phylogenetic information can help infer the MC network. The main reason is that the phylogenetic information can restrict the searching path in the solution space, hence, it is more likely to find a solution that represents the underlying true structure. 

On the contrary, other methods fail to find a good clustering of microbes. They achieve similar performance as the random model which means that the inferred MC networks do not contain enough correct information of how microbes associate with each other.

\section{Discussion and Future Work}\label{discussion}
Advances of high-throughput sequencing techniques has enabled researchers to gather metagenomic data from different environment and human niches. The available experimental data can be used to extract knowledge on how microbes interact with each other and form functional communities; these microbial functional communities are potentially related to certain microbiome induced disease. As a consequence, we propose an automated framework that can accurately identify microbial communities. We show that by incorporating phylogenetic information, our proposed PGLasso outperforms other methods in terms of ARI. 

Although we have achieved significant improvements, there are several challenges to be addressed. First, we only consider healthy samples from the HMP project; therefore, the found microbial communities may not be representative enough when applying to microbiome related disease. For example, inflammatory bowel disease (IBD) and type 2 diabetes (T2D) may show completely different microbial community structures. As a consequence, we need to conduct further research on disease related datasets. Second, our inferred microbial association may need further experiments to verify or require other automated pipeline to extract knowledge from existing published literature. These challenges are left as future work.

\end{document}